\documentclass[fleqn,10pt]{wlscirep}
\usepackage[utf8]{inputenc}
\usepackage[T1]{fontenc}
\title{Continuous-wave second-harmonic generation in the far-UVC pumped by a blue laser diode}

\author[1,2,3,*,+]{Eric~J.~Stanton}
\author[4,*,+]{Peter~Tønning}
\author[4,5]{Emil~Z.~Ulsig}
\author[4]{Stig~Calmar}
\author[1]{Maiya~A.~Bourland}
\author[4,5]{Simon~T.~Thomsen}
\author[4,5]{Kevin~B.~Gravesen}
\author[4]{Peter~Johansen}
\author[4,5]{Nicolas~Volet}

\affil[1]{EMode Photonix, Boulder, Colorado, USA}
\affil[2]{Associate of the National Institute of Standards and Technology, Boulder, Colorado, USA}
\affil[3]{Department of Physics, University of Colorado, Boulder, Colorado, USA}
\affil[4]{UVL A/S, Aarhus, Denmark}
\affil[5]{Department of Electrical and Computer Engineering, Aarhus University, Aarhus, Denmark}

\affil[*]{eric@emodephotonix.com}

\affil[+]{These authors contributed equally to this work.}

\begin{abstract}
Far-UVC light in the wavelength range of 200--230~nm has attracted renewed interest because of its safety for human exposure and effectiveness in inactivating pathogens.
Here we present a compact solid-state far-UVC laser source based on second-harmonic generation (SHG) using a low-cost commercially-available blue laser diode pump.
Leveraging the high intensity of light in a nanophotonic waveguide and heterogeneous integration, our approach achieves Cherenkov phase-matching across a bonded interface consisting of a silicon nitride (SiN) waveguide and a beta barium borate (BBO) nonlinear crystal.
Through systematic investigations of waveguide dimensions and pump power, we analyze the dependencies of Cherenkov emission angle, conversion efficiency, and output power.
Experimental results confirm the feasibility of generating far-UVC, paving the way for mass production in a compact form factor.
This solid-state far-UVC laser source shows significant potential for applications in human-safe disinfection, non-line-of-sight free-space communication, and deep-UV Raman spectroscopy.
\end{abstract}
\begin{document}

\flushbottom
\maketitle
\thispagestyle{empty}

\section*{Introduction}

Over the past decade, extensive research has substantiated the safety of human exposure to far-UVC light (wavelengths from 200~nm to 230~nm) and its effectiveness in inactivating diverse pathogens, including human coronaviruses \cite{buonanno_2017, ponnaiya_2018,buonanno_2020,kitagawa_2021,ma_2021,ma_2021uv,glaab_2021,sugihara_2022,blatchley_2023,welch_2023}.
These compelling findings underscore the immense potential of far-UVC sources in curtailing airborne viral transmission, thereby playing a crucial role in the containment of diseases like COVID-19 and averting future pandemics.
Additionally, far-UVC may be instrumental to improve yields in farming and pharmaceutical industries \cite{singh_2021, subedi_2021, ramos_2020}.
Previous studies have primarily investigated the disinfection and interactions of food, livestock, and pathogens with longer wavelengths (250--270 nm), which is unsafe for direct human exposure.
However, employing human-safe far-UVC in these applications offers a notable advantage.
By helping to eliminate contaminants, far-UVC could bring about additional societal benefits by enhancing the quality of goods in food and drug supply chains and curbing disease outbreaks that stem from these industries \cite{grida_2020,rowan_2020,abdolazimi_2023,miranda_2019,rizou_2020}.
Beyond disinfection, compact far-UVC light sources 
can simplify free-space communication systems by eliminating the need for pointing accuracy and stability
\cite{Cai_2021}.
The reduced solar background noise in this spectrum and a broad acceptance angle may enable non-line-of-sight links, potentially reducing both the cost and complexity of the overall system.
Furthermore, far-UVC lasers are beneficial for Raman spectroscopy \cite{Hara_2022}, as they provide greater detail and higher quality information about a sample than lasers at longer wavelength.
The higher photon energy increases the Raman scattering efficiency and reduces the fluorescence background in numerous materials.

Achieving widespread and strategic deployment of far-UVC sources demands substantial technological advancements in the field \cite{Kneissl_2019}.
The current leading commercial technology for far-UVC disinfection is the krypton chloride (KrCl) excimer lamp, emitting $\sim$150~mW with a spectral peak at 222~nm \cite{Eadie_2022}.
These devices use high-voltage ($>$1 kV) radio-frequency discharges resulting in a large power consumption.
Inherently, their footprint is large (cm size), and they require a filter to eliminate dangerous wavelengths longer than 230~nm.
While the KrCl emission wavelength cannot be tuned, slightly shorter wavelengths are anticipated to be safer and more effective in pathogen elimination  \cite{hessling_2021}.
Since the intensity of the lamp depends on the excimer volume ($\sim$1~cm$^3$), the emission is difficult to control and inefficiently coupled to an optical fiber.

An attractive solution to reduce size, weight, operating power, and cost is direct emission of far-UVC from a semiconductor material.
The performance of light-emitting diodes (LEDs) are progressing \cite{Knauer_2023}.
However, for laser diodes, the shortest emission wavelength to date is $>$270~nm \cite{zhang_2022}, and fundamental obstacles remain that could potentially limit lifetime, efficiency, and output power.

\begin{figure}[tb]
\centering
\includegraphics[width=\textwidth]{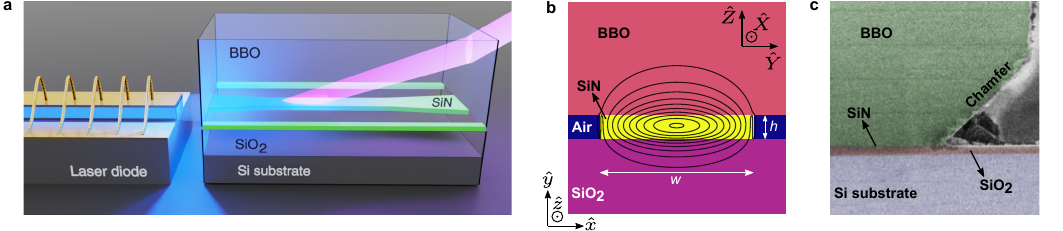}
\caption{
\textbf{a},~Schematic of chip-to-chip edge coupling concept between a laser diode and the frequency converter showing the far-UVC generated at an angle above the waveguide core.
\textbf{b},~Waveguide cross-section schematic showing the BBO nonlinear crystal bonded to the SiN waveguide core of width $w$ and thickness $h$.
The Si substrate (not shown in this cross-section) is located below the SiO$_2$ lower cladding.
Geometrical axes $\hat{x}$, $\hat{y}$ and $\hat{z}$ are shown, as well as crystal axes $\hat{X}$, $\hat{Y}$ and $\hat{Z}$ of BBO.
Contour lines show the simulated $\hat{x}$-polarized electric field profile of the fundamental TE mode at a wavelength of 445~nm.
The SiN core guides the mode with a confinement of 58~\%, and a significant portion of the evanescent field (26~\%) overlaps with the BBO.
\textbf{c},~Scanning electron micrograph showing the BBO crystal directly bonded to the SiN waveguide, with the SiO$_2$ lower cladding and the Si substrate.
Edges of the BBO sample are chamfered for mechanical stability.
}
\label{fig:waveguide}
\end{figure}

We propose to convert light from a continuous-wave (CW) blue laser diode to the far-UVC using second-harmonic generation (SHG) in a waveguide, as depicted in Fig.~\ref{fig:waveguide}a.
This approach leverages decades of research on GaN emitters for lighting \cite{pimputkar_2009,behringer_2020} and incorporates recent advances in nanophotonic waveguide manufacturing \cite{blumenthal_2018,chanana_2022}.
The benefits of this approach include a low-voltage power supply, the low-cost and high-availability of blue laser diodes, wavelength flexibility, and the potential for low-cost fabrication with wafer-scale semiconductor manufacturing.

Beta barium borate (BBO) is commonly used for second-harmonic generation to the UV \cite{eimerl_1987}.
This is thanks to its broad transparency range, large nonlinearity and a shorter phase-matching cut-off wavelength compared to other nonlinear crystals \cite{Kang_2022,Mutailipu_2023}.
A relatively compact demonstration with an external-cavity diode laser (ECDL) pump at 445~nm achieved far-UVC using bulk BBO, though with low conversion efficiency ($\sim$0.01~\% for 1.4~W pump) \cite{ruhnke_2018}.
Recent advancements using resonant cavities showed improved efficiency with 13~\% conversion from a 200~mW pump and 34~\% conversion using cascaded SHG from a 1.6~W pump, both producing signals at $\sim$229~nm \cite{Foster_2022, kaneda_2016}.
However, these systems require complex bulk optics for implementation.

Another approach for frequency conversion to the far-UVC involves a pulsed pump source for optical parametric oscillation and SHG.
Demonstrations achieved 16.2~\% efficiency from a 1.2~mJ pump, generating pulses at 222~nm wavelength \cite{Luo_2023}.
Compressed pulses were also used in He-filled hollow capillary fibers, achieving 3.6~\% conversion from 0.2~mJ pulses \cite{brahms_2023}.
Pulse-pumping offers higher conversion efficiency but requires more complex and larger pump laser systems.

SHG in BBO waveguides was demonstrated at 266~nm wavelength with negligible walk-off and achieved 0.05~\% conversion efficiency from a 670~mW pump \cite{degl_2008}.
However, phase-matching was limited to the bulk phase-matching angle, not utilizing the dispersion properties of the waveguide.

Our approach uses Cherenkov phase-matching \cite{suhara_1993} across a bonded interface, showcased in Fig.~\ref{fig:waveguide}a,b, with a scanning electron micrograph in Fig.~\ref{fig:waveguide}c.
Blue pump light propagates in a waveguide, creating an evanescent field in a nonlinear crystal directly bonded to the waveguide core.
This heterogeneous platform enables far-UVC generation without any interaction of the signal light with the waveguide core that guides the pump.
Combining the precision and scalability of Si nanophotonics with the high nonlinearity and low loss of a bulk crystal is the key to achieving high conversion efficiency and stable operation.

\section*{Results}

This work showcases experimental demonstrations of a heterogeneous photonic chip generating far-UVC with SHG from a blue diode laser.
It is complemented by novel simulation results for the frequency converter chip design. 

\begin{figure}[tb]
\centering
\includegraphics[width=\textwidth]{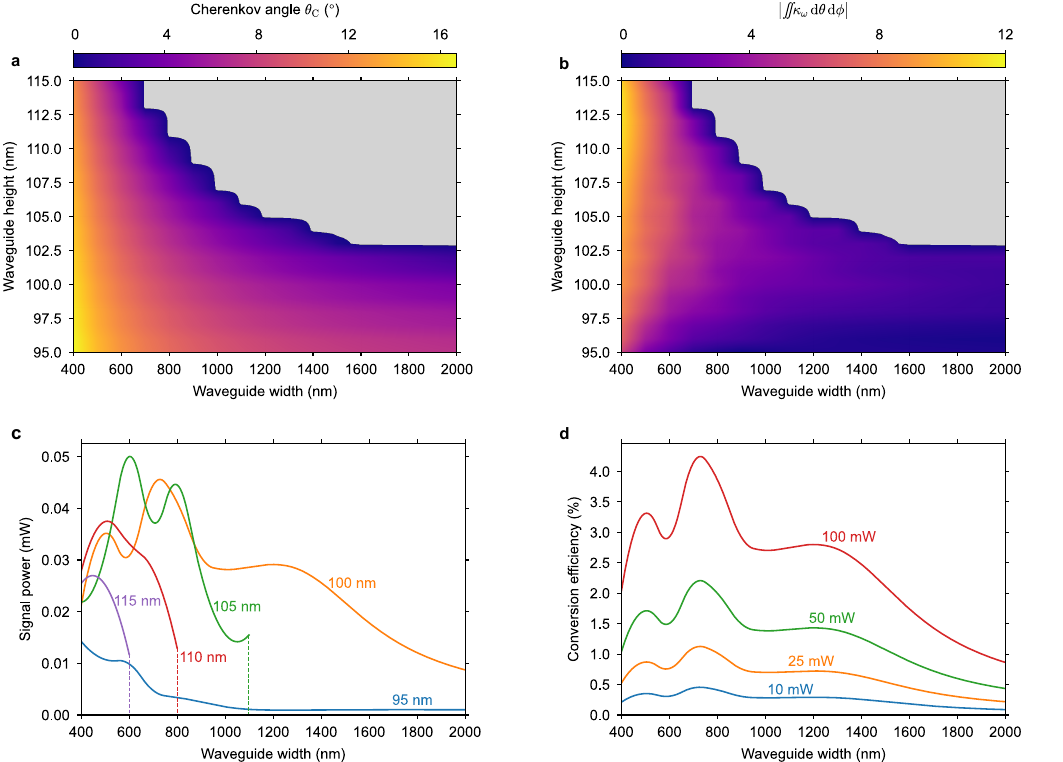}
\caption{
\textbf{a},~Cherenkov angle simulated for the BBO/SiN hybrid waveguide for various thicknesses and widths of the SiN waveguide core.
\textbf{b},~Cherenkov nonlinear coupling parameter in a BBO/SiN waveguide as a function of the SiN core height and width. \textbf{c},~Trends of the signal power with waveguide width for various waveguide heights as annotated on the plot for a pump power of 50~mW.
\textbf{d},~Trends of the conversion efficiency with waveguide width for various pump powers as annotated on the plot for a waveguide height of 100~nm.}
\label{fig:Cherenkov}
\end{figure}

\subsection*{Numerical simulation and analysis}

Phase-matching is found for various SiN waveguide heights and widths (Fig.~\ref{fig:Cherenkov}a), with predicted Cherenkov angle values ranging from zero to $\sim$16\textdegree{}.
In regions with large waveguide width and height, phase-matching collapses.
The Cherenkov angle is close to zero along this boundary and generally increases for smaller widths and heights away from the boundary cut-off condition.
While having the Cherenkov angle close to zero facilitates a low-angle emission at the BBO output facet, this work focuses on optimizing the conversion efficiency, irrespective of the Cherenkov angle.

\begin{figure}[t]
\centering
\includegraphics[width=\textwidth]{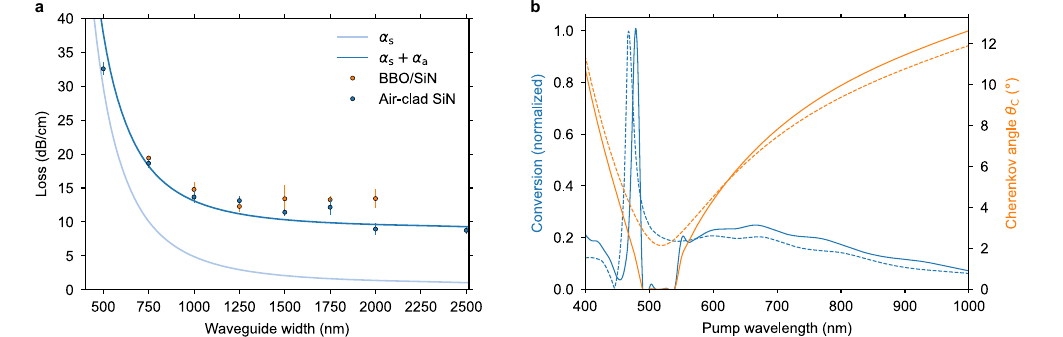}
\caption{
\textbf{a},~Scattering loss ($\alpha_\mathrm{s}$) of TE-polarized modes in waveguides designed for Cherenkov phase-matching with 98~nm thick SiN, a sidewall roughness $\sigma_\mathrm{side} = 7~\mathrm{nm}$ RMS, and a top surface roughness $\sigma_\mathrm{surface} = 0.25~\mathrm{nm}$ RMS.
The absorption loss $\alpha_\mathrm{a}$ is calculated from the material absorption loss of SiN scaled by the confinement factor, where $\alpha_\mathrm{SiN}$ = 14~dB/cm. Measured propagation losses for the BBO/SiN and the air-clad SiN waveguides are plotted with data points, and error bars indicate the standard deviations of the measurement uncertainties.
\textbf{b},~Normalized conversion efficiency for SHG simulated across a broad range of pump wavelengths, plotted in blue on the left axis.
The corresponding Cherenkov angle is plotted in orange on the right axis.
Solid lines represent a waveguide with a 100~nm height and a 1500~nm width, which is the geometry optimized for SHG from a 445~nm wavelength pump.
Dashed lines represent a waveguide with a 96~nm height and a 2000~nm width, showcasing a finite frequency conversion across the entire pump wavelength range of 400~nm to 1000~nm.
}
\label{fig:sim_trends}
\end{figure}

The strength of the nonlinear interaction, independent of propagation loss, is observed by $\iint \kappa_\omega \, \mathrm{d}\theta \, \mathrm{d}\phi$ (as described in \hyperref[Methods]{Methods}), and is plotted in Fig.~\ref{fig:Cherenkov}b.
The results indicate that narrower and thicker waveguides provide a stronger frequency conversion.
However, the realized frequency conversion takes into account the losses, resulting in an optimal point for the SiN geometry, as propagation losses are inversely proportional to the waveguide width.
Figure~\ref{fig:sim_trends}a shows the simulated propagation loss contribution from interfacial scattering and the sum of scattering and absorption losses in the SiN.
Notably, measurements of the propagation loss align well with the predicted total loss values for the measured widths ranging from 500~nm to 2500~nm.
From these data, the signal power and the conversion efficiency are predicted for selected waveguide heights (95~nm to 115~nm).
Figure~\ref{fig:Cherenkov}c shows the signal power trend with waveguide width for each height, with 100~nm being the preferred height due to its large peak value and broad phase-matching range.
In Fig.~\ref{fig:Cherenkov}d, the conversion efficiency is presented as a percentage of the pump power for this height at various pump powers.
The on-chip pump power of 100~mW achieves a maximum efficiency of $\sim$4~\%.

\begin{figure}[tb]
\centering
\includegraphics[width=\textwidth]{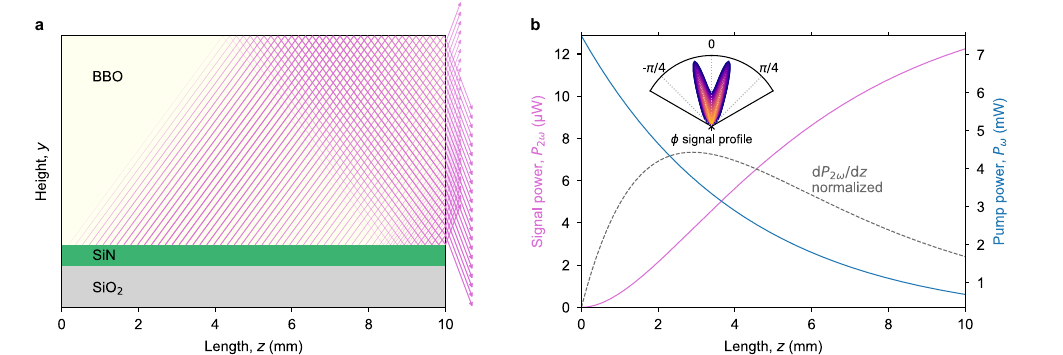}
\caption{
\textbf{a},~Ray diagram of the signal generation in the BBO along the SiN waveguide.
The ray width and transparency is scaled by the relative generated signal power per unit length, and the output at the right facet accounts for refraction.
The SiN and SiO$_2$ layers are shown thicker than the actual layers for illustrative purposes. \textbf{b},~The signal power $P_{2\omega}$ is generated along the propagation length $z$ as the pump power $P_{\omega}$ depletes. 
The normalized differential of the signal power with the length $z$ is shown with a dashed gray line, corresponding to the signal generation rays in \textbf{a}.
The dependence of $P_{2\omega}$ on the azimuthal angle $\phi$ is shown in the inset.
}
\label{fig:z_conversion}
\end{figure}

Figure~\ref{fig:z_conversion}a illustrates the signal power generated along the propagation of the pump waveguide, while Fig.~\ref{fig:z_conversion}b plots the signal and pump powers along $z$.
The initial section of the waveguide experiences faster pump to signal conversion (indicated by the dashed grey line in Fig.~\ref{fig:z_conversion}b) due to the attenuation of pump light by propagation loss.
Additionally, the ray diagram in Fig.~\ref{fig:z_conversion}a illustrates the output beam realization during the experiment.
The larger-than-anticipated Cherenkov angle results in the signal reflecting from the top of the chip, leading to two emission beams, both downward and upward.
This unexpected behavior is partly due to fabrication tolerance on the SiN layer thickness, resulting in a 98~nm thick layer instead of the targeted 100~nm.
Furthermore, the BBO chip thickness was reduced due to an additional polishing step on the top surface to aid the coupling procedure with a microscope view.
The azimuthal radiation profile, as shown in the inset of Fig.~\ref{fig:z_conversion}b, exhibits a local minimum at $\phi = 0$ attributed to the aspect ratio of the SiN waveguide.
This effect can be mitigated by engineering a waveguide with a narrower width and thicker height.

While this device was designed for optimal operation with a 445~nm pump wavelength, other SiN geometries can support extremely broad-band phase-matching. 
For a waveguide thickness of 96~nm and a width of 2000~nm, Cherenkov phase-matching spans a continuous range of pump wavelengths from 400~nm to 1000~nm (Fig.~\ref{fig:sim_trends}b).
While the trend ignores variation in propagation loss across this spectrum, it underscores the robustness of the Cherenkov approach, requiring no fine-tuning of refractive indices for phase-matching.
Moreover, this discovery opens possibilities for the device's realization in other applications involving broadband frequency conversion.

\subsection*{Passive transmission}

Firstly, we characterize devices comprising air-clad passive SiN waveguides without the bonded BBO to assess the relative contributions of scattering and absorption effects on propagation loss.
The transmission of blue laser light is measured from straight waveguides to determine propagation and coupling losses.
These results, along with the predicted trends, are plotted in Fig.~\ref{fig:sim_trends}a.

Subsequently, we measure blue light transmission with a BBO crystal bonded to the SiN waveguide layer.
The TE-polarized light at 445~nm wavelength from a 2.0~\textmu{}m wide waveguide exhibits a propagation loss of $\alpha_\mathrm{p}$ = $13.4 \pm 1.3$~dB/cm, while the coupling loss to a lensed fiber is 11.0~dB.
The presence of observable grain residues around the waveguide facet, arising from hybrid facet polishing, negatively impacts the coupling loss.
This is evidenced by the trend of lower coupling losses for wider waveguides, where the light has less interaction with the SiN sidewall interface.
However, wider waveguides result in increased relative coupling to higher-order modes, reducing the portion of light contributing to frequency conversion.
To balance the effects of insertion loss, propagation loss, and single-mode propagation, there is an optimal waveguide width designed for the fundamental TE mode phase-matching.
This optimal width is experimentally found in waveguides ranging from 1.0~\textmu{}m and 2.0~\textmu{}m in the following section.

\subsection*{SHG to the far-UVC}

\begin{figure}[tb]
\centering
\includegraphics[width=\textwidth]{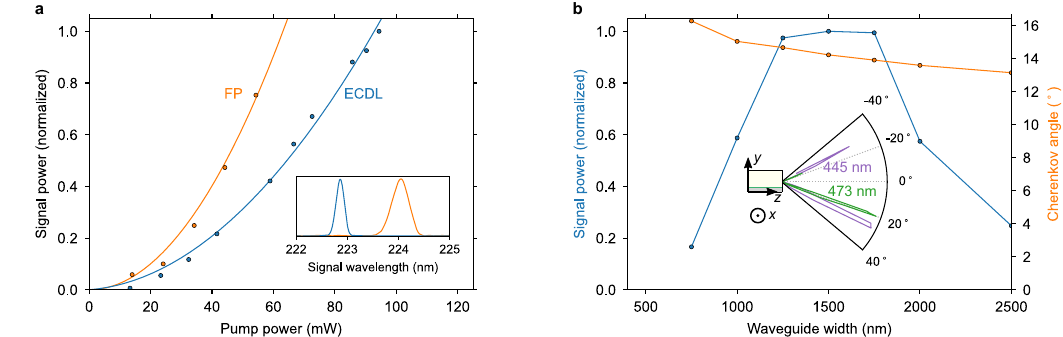}
\caption{
\textbf{a},~Far-UVC power quadratic dependence on the off-chip pump power.
The inset shows the far-UVC signal from an ECDL laser and an FP diode laser.
\textbf{b},~Waveguide width dependence on the far-UVC output intensity and the Cherenkov angle.
The inset shows the angular dependence on the far-UVC output for pump wavelengths of 445~nm and 473~nm from a 1500~nm wide waveguide.
}
\label{fig:output}
\end{figure}

Far-UVC generation is measured with a blue laser diode pump and detecting the emitted light at various output angles from the BBO facet.
Figure~\ref{fig:output}a illustrates the output power dependence on the pump power using both ECDL and Fabry-Perot (FP) pump lasers at 445~nm, as shown in Fig.~\ref{fig:output}a (see \hyperref[Methods]{Methods} for details on laser sources used).
Quadratic fits to both types of pump lasers result in coefficients of determination exceeding 98~\%.
Interestingly, the conversion efficiency from the FP laser appears unaffected by the expansion of lasing frequency modes when increasing the drive current.
The observed difference in conversion efficiency is attributed to differences in the coupling loss of approximately 2.2~dB.

With a fixed pump power from a single-frequency laser at 445~nm wavelength, we measured the SHG output intensity for various emission angles in the subset of Fig.~\ref{fig:output}b.
SHG light generated at the beginning of the waveguide can reflect at the top of the BBO chip and exit the BBO facet at the negative of the Cherenkov phase-matched angle, with refraction at the exit facet.
The same measurement was repeated with input wavelengths of 405~nm and 473~nm, corresponding to larger and smaller Cherenkov angles as predicted by simulations in Fig.~\ref{fig:sim_trends}b.

By changing the waveguide width, the effective index of the pump mode is varied, thereby modifying the Cherenkov angle and the conversion efficiency.
In Fig.~\ref{fig:output}b the angle of maximum emission and the relative power at a fixed pump power is recorded as a function of the waveguide width.
Although the far-UVC power is lower for waveguide widths of 750~nm and 1000~nm compared to wider widths, we expect this discrepancy to be attributed to facet coupling efficiency rather than a trend in the conversion efficiency.
The trend in the Cherenkov angle predicted by simulations in Fig.~\ref{fig:Cherenkov} is confirmed with an additional offset.

\begin{figure}[tb]
\centering
\includegraphics[width=\textwidth]{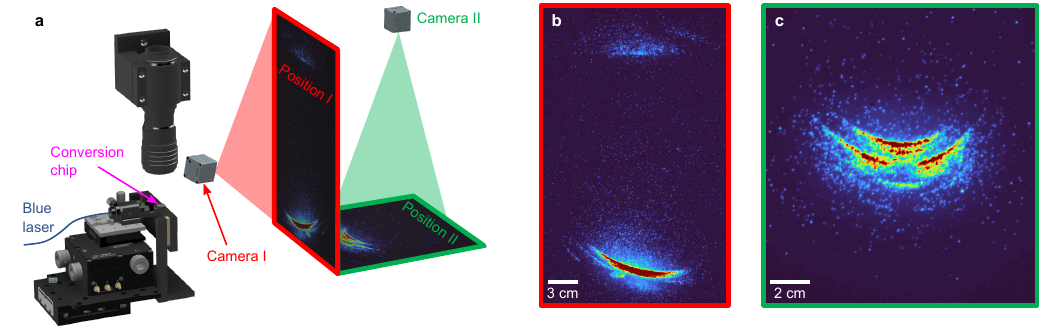}
\caption{
\textbf{a},~Setup for capturing images of the far-field showing the placement and orientation of both camera and projection screen.
The microscope shown above the conversion chip aids the alignment process of the lensed fiber to the chip.
\textbf{b,c},~Images of the far-UVC reflected off the projection screen, with the upper emission lobe less defined than the lower due to the multiple reflections through the BBO.
A considerable spread in the azimuthal angle is observed as phase-matching is achieved with a range of Cherenkov angles.
}
\label{fig:UVcamera}
\end{figure}

Beyond the single axis distribution obtained from the goniometer, the far-field is further explored using a UV camera.
A schematic of the setup with the camera placement and projection screens (described in \hyperref[Methods]{Methods}) is displayed in Fig.~\ref{fig:UVcamera} along with images captured of the screen positioned $\sim$20~cm from the chip.
From Fig.~\ref{fig:UVcamera} both upward and downward emission lobes are evident, with the lower emission significantly more intense, as explained from the geometrical considerations in Fig.~\ref{fig:z_conversion}. 
We also observe that Cherenkov phase-matching is achieved for range of azimuthal angles ($\phi$).

\subsection*{Wavelength tuning and SFG}

\begin{figure}[tb]
\centering
\includegraphics[width=\textwidth]{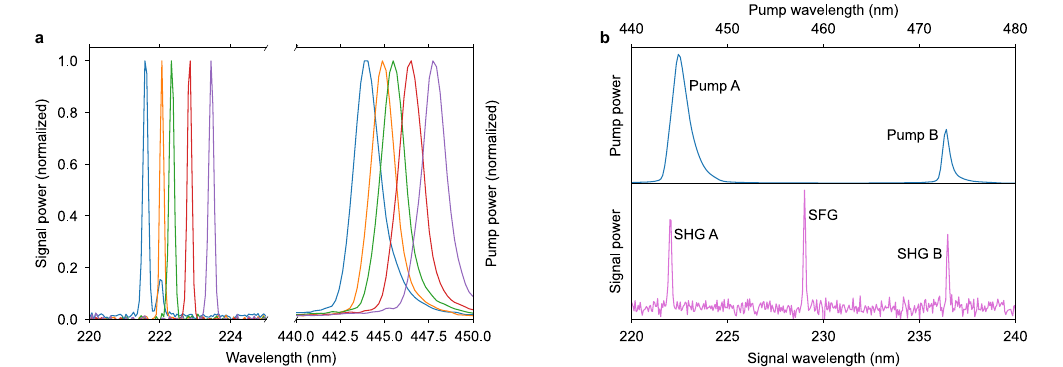}
\caption{
\textbf{a},~Tuning the single-frequency pump laser produces a continuous sweep of far-UVC output wavelengths.
\textbf{b},~Using two pump lasers at respectively 445~nm and 473~nm, sum-frequency generation is demonstrated at 229.5~nm.
The spectra, showing the normalized detected power, are recorded with two different optical spectrometers, the UV spectrometer having significantly higher resolution.
As such the broader spectral appearance of the pump lasers is not their true spectral width as all spectra are under-sampled and merely reflect the resolution of the spectrometer.
}
\label{fig:WL_tuning}
\end{figure}

As shown in Fig.~\ref{fig:output}a, the frequency converter exhibits tolerance to significant wavelength shifts, attributed to the robustness of Cherenkov phase-matching.
This characteristic enables tunability, a valuable feature for several key applications in the far-UVC domain.
Figure~\ref{fig:WL_tuning}a demonstrates the direct mapping from wavelength tuning of the blue pump laser to the corresponding SHG signal across the tuning range of the stabilized diode laser.

The spectral tolerance of the frequency converter is further explored using different pump lasers to extend beyond the tuning range of the 445~nm laser.
While the current design of the frequency converter chip is not optimized for pump wavelengths below 440~nm due to increased propagation loss in the SiN waveguides and lower conversion efficiency, slightly longer wavelengths support even higher conversion efficiency.
This is demonstrated with a pump laser at 473~nm wavelength, as shown in Fig.~\ref{fig:output}b, where the Cherenkov angle is lower for the 473~nm pump as predicted in simulations, and the conversion efficiency is higher.

A far-UVC signal is also detected for a pump laser at 405~nm, as described in \hyperref[Methods]{Methods}.
However, due to higher losses indicating lower conversion efficiency, the output signal is very low, and thus, no spectrum was recorded at this wavelength.
The output emission at 202.5~nm is observed at a steeper output angle compared to the longer far-UVC wavelengths.
Most of the power lies in the upper emission lobe, as explained by Fig. \ref{fig:z_conversion}a, where greater angles lead to double reflection of the signal generated in the first part of the chip.

Regardless, this result is significant because the pump wavelength of 405~nm is below the range for conventional phase-matching of SHG in bulk BBO \cite{Kang_2022,Mutailipu_2023}.
In contrast, Cherenkov phase-matching can support efficient SHG well below 205~nm, which is promising for enabling unanticipated applications at wavelengths below 200~nm in the vacuum-UV.

By combining two pump lasers at 445~nm and 473~nm wavelengths, we can further explore the tolerance of phase-matching through sum-frequency generation (SFG). 
Both SHG and SFG are observed in Fig.~\ref{fig:WL_tuning}b when each laser is coupled to the same waveguide on the frequency converter chip.
Owing to the broadband phase-matching, a single waveguide simultaneously supports SHG at 445~nm and 473~nm, with slightly different Cherenkov angles.
Additionally, a SFG signal is produced with a wavelength of 229.5~nm.

\section*{Discussion}

In conclusion, we have successfully demonstrated a solid-state far-UVC laser source using a chip-scale pump laser.
This technology offers significant potential for mass production, benefiting from the extensive development of blue laser diodes.
The broad phase-matching bandwidth is pivotal in achieving high-yield production with robust operation, ensuring the device's resilience against fabrication tolerances and environmental variations.
As a result, pump sources like mass-produced FP laser diodes can be readily employed, facilitating the development process.
The frequency converter utilizes standard materials and minimizes the material processing of BBO.
This combination allows for a substantial reduction in size and price, thereby enabling the targeted high-volume application of disinfection.

Further, a compact far-UVC laser could drive other applications like non-line-of-sight communication, leveraging the drastic increase of scattering at shorter wavelengths \cite{Cai_2021} or Raman spectroscopy where excitation at wavelengths shorter than 250~nm alleviates autofluorescence \cite{Hara_2022}.
Utilizing SFG, the spectral acceptance of pump sources enables compact realization of laser sources at a broad range of wavelengths.
For example, combinations of compact diode lasers could be used to generate laser emission spanning the UV and visible spectra.
We also envision that a frequency-comb pump source could be translated to the UV with this frequency converter device because of the broadband phase-matching.

The main limiting factor in the presented work is the loss associated with input coupling and propagation through the SiN waveguide.
As shown in Fig. \ref{fig:output}, an increase in on-chip pump power has significant impact on the signal power.
Previous works demonstrating SiN waveguides with similar geometries show lower propagation losses of 5--7~dB/cm \cite{sorace_2019,Chauhan_2022,smith_2023}.
Other means of improvements lie in the material platform.
Some evidence suggest lower loss from PECVD SiN \cite{sorace_2019}.
Also, tantala could provide lower material absorption while maintaining a similar refractive index to SiN \cite{spektor_2023}.
Gradual improvements in blue laser power densities are still seen from manufacturers, and pulsed operation could generate a higher power signal relevant for disinfection.
The theory and architecture presented in this study provide multiple avenues for enhancing conversion efficiency, thereby increasing the power output of this compact solid-state far-UVC laser.
Together, these advancements pave the way for far-UVC laser applications in non-line-of-sight free-space communication, Raman spectroscopy, and disinfection.

\section*{Methods}\label{Methods}

\subsection*{Design}

SiN is used for the waveguide core layer to guide the blue pump light.
BBO is used as the nonlinear crystal because of its large second-order nonlinearity, low loss in the far-UVC, and refractive indices compatible with Cherenkov phase-matching to SiN waveguides \cite{eimerl_1987, tamovsauskas_2021}.
Figure~\ref{fig:waveguide}b shows a schematic to define the waveguide width ($w$), height ($h$), the layer compositions, and the crystal axes relative to the device coordinate system.
The BBO crystal is orientated to align $d_{16}$ of the second-order nonlinear tensor to the TE-like pump mode in the SiN \cite{volet_2021}.
A contour profile of the $\hat{x}$-polarized electric field of the pump at 445~nm wavelength overlays the schematic.
It has high confinement in the SiN core and a substantial overlap of the evanescent field with the BBO.

The first step in the design is to investigate geometries of SiN waveguides that support Cherenkov phase-matching to radiation modes in BBO bonded directly to SiN.
With the propagation in the waveguide defined as the $\hat{z}$-direction, Cherenkov phase-matching occurs when the signal light propagates in a direction slightly elevated from the $\hat{z}$-direction towards the $\hat{y}$-direction, such that the $\hat{z}$-component of the phase velocity matches the pump phase velocity.
The Cherenkov angle ($\theta_\mathrm{C}$) is the elevation angle $\theta$ of the signal wavevector relative to the SiN waveguide for the azimuthal angle of $\phi = 0$.
This phase-matching condition occurs when:
\begin{equation}
    n_{\omega} 
    = n_{2\omega} 
    \cos \!\left( \theta_\mathrm{C} \right) \,,
\end{equation}
where $n_{\omega}$ is the effective refractive index of the guided pump mode and $n_{2\omega}$ is the refractive index the radiated signal light experiences in the nonlinear material.
It is illustrated in Fig.~\ref{fig:waveguide}a and Fig.~\ref{fig:z_conversion}a, and measured in Fig.~\ref{fig:output}a.
Note that Cherenkov phase-matching can occur only when $n_{2\omega} \geqslant n_{\omega}$.
This requirement dictates the materials that can be used for the core, cladding, and nonlinear regions of the Cherenkov SHG waveguide.
It also creates a regime of operation for certain combinations of waveguide widths and heights such that $n_{\omega}$ is engineered to be smaller than $n_{2\omega}$.

The derivation for the Cherenkov SHG conversion efficiency follows the method proposed in \cite{suhara_1993}, with some modifications to include propagation losses. 
The spatially dependent electric and magnetic fields (of either the pump or the signal) are assumed to take the form of:
\begin{subequations}
\begin{align}
\vec{\mathcal{E}}(x,y,z) 
&= 
e^{i k z} \,
\mathcal{Z} (z)
\vec{E}(x,y) , \\
\vec{\mathcal{H}}(x,y,z) 
&= 
e^{i k z} \,
\mathcal{Z} (z)
\vec{H}(x,y) ,
\end{align}
\end{subequations}
where $\vec{E}$ and $\vec{H}$ are the electric and magnetic field components of the mode profile, assumed to be independent of $z$, and $\mathcal{Z}$ is a unitless complex function of $z$ that accounts for nonlinear mode coupling.
The term $e^{i k z}$ with $k = \beta + i \alpha/2$ describes the propagation in the $\hat{z}$-direction through the associated component of the wavevector $\beta$ and the power loss coefficient $\alpha$.
The pump mode profiles are calculated by a numerical simulation method \cite{emode}.
The signal consists of a continuum of radiation modes that are best expressed analytically \cite{Papakonstantinou_2009}. 
The approach taken by \cite{suhara_1993} for embedded channel waveguides is the approximation we use for this waveguide geometry. The $\hat{x}$-polarized electric field profile for TE-like radiation modes is calculated for an azimuthal range of: $-\pi/2 < \phi < \pi/2$.

Nonlinear coupling can be modeled with the following equation \cite{Stegeman_1985}:
\begin{equation}
\partial_z \mathcal{Z}_\nu 
= i \frac{\omega_\nu}{4 Q_\nu} 
e^{-i k_\nu z} \iint_A \vec{\mathcal{P}}_{\nu}^\mathrm{NL} \cdot \vec{E}_{\nu}^* 
\,\mathrm{d}x \,\mathrm{d}y ,
\end{equation}
with a normalization parameter
(in units of power):
\begin{equation}
Q_{\nu} = \frac{1}{2}
\iint_A \Re\left[ 
\left( \vec{E}_{\nu} 
\times \vec{H}_{\nu}^* \right)
\cdot \hat{z}
\right] \mathrm{d}x \, \mathrm{d}y ,
\end{equation}
where the integration is performed over a finite cross-sectional area $A$ that completely encompasses the pump field.
The nonlinear electrical polarizations are defined as $\vec{\mathcal{P}}_\mathrm{\omega}^\mathrm{NL} = 2 \epsilon_0 d \vec{v}_{\omega}$ for the pump and $\vec{\mathcal{P}}_\mathrm{2\omega}^\mathrm{NL} = \epsilon_0 d \vec{v}_{2\omega}$ for the signal, $\epsilon_0$ is the vacuum permittivity, $d$ is the tensor representing the second-order nonlinearity of the material, and $\vec{v}$ is the corresponding complex vector of pump or signal electric field components assuming Kleinman symmetry.
Next, the nonlinear coupling coefficients are calculated as a function of the elevation angle $\theta$ and of the azimuthal angle $\phi$ by:
\begin{subequations}
\begin{align}
\kappa_{2\omega} 
&= \frac{\omega \epsilon_0}{2 Q_{2\omega}} \iint_A (d\vec{v}_{2\omega}) \cdot \vec{E}_{2\omega}^* \, \mathrm{d}x \, \mathrm{d}y ,
\\
\kappa_{\omega} 
&= \frac{\omega \epsilon_0}{2 Q_{\omega}} \iint_A (d\vec{v}_{\omega}) \cdot \vec{E}_{\omega}^* \, \mathrm{d}x \, \mathrm{d}y .
\end{align}
\end{subequations} 
Now the coupled-amplitude equations can be expressed as:
\begin{subequations}
\begin{align}
\partial_z \mathcal{Z}_{2\omega} 
&= i \mathcal{Z}_{\omega}^2 \kappa_{2\omega} 
e^{-i z \Delta k} ,
\\
\partial_z \mathcal{Z}_{\omega} 
&= i \mathcal{Z}_{\omega}^* \iint \mathcal{Z}_{2\omega} \kappa_{\omega} e^{(i\Delta k - \alpha_\omega)z} \,\mathrm{d}\theta \,\mathrm{d}\phi  ,
\end{align}
\label{eq:CA}
\end{subequations}
where $\Delta k = k_{2\omega} - 2 k_\omega$.
Note that the signal amplitude $\mathcal{Z}_{2\omega}$ is a function of $\theta$ and $\phi$.
The coupled-amplitude equations in Eq.~\ref{eq:CA} are solved numerically to find the signal output power $P_{2\omega}(z) = Q_{2\omega} e^{-\alpha_{2\omega} z} \left| \iint \mathcal{Z}_{2\omega} \, \mathrm{d} \theta \, \mathrm{d}\phi \right|^2$ in the radiated fields and the pump power $P_{\omega}(z) = Q_{\omega} e^{-\alpha_{\omega} z} \left| \mathcal{Z_{\omega}} \right|^2$ in the guided pump mode.
The input pump power $P_{\omega,0}$ initializes the pump amplitude as $\left| \mathcal{Z}_{\omega}(0) \right|^2 = P_{\omega,0} / Q_{\omega}$, and the absence of an initial signal sets $\mathcal{Z}_{2\omega}(0) = 0$.
Note that the radiation fields and $\kappa$ must be defined over the range of $\phi$ and a range of $\theta$ large enough to include all contributions.

\subsection*{Fabrication}

Device fabrication starts with an oxidized Si wafer, on which a target of 100~nm SiN is formed by low-pressure chemical vapor deposition.
The actual thickness of this layer is measured as 98~nm.
Waveguides are patterned with deep-UV lithography and etched with inductively-coupled plasma reactive ion etching.
Chips are singulated by dicing to a propagation length of 10~mm. 
BBO crystals are then grown and cut along the $\hat{X}$ crystal axis with a similar propagation length. The $\hat{Z}$ and $\hat{X}$ axes are polished. 
The SiN and BBO chips are directly bonded together after O$_2$ plasma activation, and the bond is strengthened with an anneal at 300~\textdegree{}C for 12~hours.
Hybrid facets of Si and BBO are formed at the input and output of the waveguide by grinding and polishing to a single planar and smooth surface using a water-free polishing solution because BBO is hygroscopic.
The facets are coated with MgF anti-reflective layers for the pump at the input side of the chip and for the signal at the output side.

\subsection*{Measurements}

For all experiments a laser is coupled to the edge facet of the chip with lensed fibers.
Polarization-maintaining (PM) fibers mounted on a 3-axis nanopositioning translation stage ensure close control of both polarization and spatial alignment.
Detection equipment on the output side of the setup varies and is described for each measurement.

The following lasers are used to pump the nonlinear processes.
A stabilized 445-nm diode laser (DLPro from Toptica) is used to pump SHG and SFG, and it is referenced as the ECDL.
This laser has a manual tuning range of $\sim$4~nm a maximum power in fiber of 95~mW and an instantaneous linewidth of 150~kHz.
A fiber-coupled 445-nm FP diode laser (PL450B from Osram, packaged by OZ Optics) is used for SHG, and has a maximum power in fiber of 55~mW and a spectral width of $\sim$1.5~nm corresponding to 10 to 15 longitudinal cavity modes.
A fiber-coupled FP diode laser emitting around 405~nm (NDV4316 from Nichia, packaged by OZ Optics) is used for SHG, emitting a power of 45~mW in fiber and a $\sim$1~nm spectral width, corresponding with 8 to 12 longitudinal cavity modes.
Finally, a stabilized free-space diode laser at 473~nm with an intantaneous linewidth $<$1~MHz (Cobolt08 from Hübner Photonics) is used for SHG and SFG.
It is coupled to a single-mode PM fiber achieving 15~mW in the fiber.

For passive transmission measurements, PM lensed fibers are used for both the input and output coupling.
Two techniques are used to discriminate the coupling loss from propagation loss. 
First the total power transmission is measured through detection on a fiber-coupled photodiode (S130VC from Thorlabs), secondly the scattered light intensity is measured along the waveguide from a microscope on top of the chip with a visible-light camera \cite{daldosso_2004}.

For characterizing the far-UVC signal from the frequency converter, the input coupling remains the same, while the output is collected using a custom designed 1D goniometer.
The goniometer allows for mounting of different sensors recording the output at any output angle, vertically with reference to the waveguide at a fixed distance of 10~cm.
For detection of the UVC signals either an optical spectrometer (Maya2000 Pro from Ocean Insight) or a photoemission detector (UVtron from Hamamatsu, with Ni electrode) is used.
The photoemission detector gives a binary detection signal with very high sensitivity, allowing for investigations outside of the main emission lobe, however this detector has a low dynamic range.
To measure the output angle $\theta$, shown in Fig.~\ref{fig:output}b, the photoemission detector is mounted on the 1D goniometer.
The limited dynamic range leads to saturation at the main emission lobe from the 445~nm pump laser.
Due to its high directionality, lack of calibration, and limited dynamic range, the photoemission detector cannot be effectively mapped to a calibrated power-meter within this range.
The spectrometer on the other hand has a linear intensity response, excellent dynamic range, but measures intensity in a small aperture of the input multi-mode fiber of 600~\textmu{}m.
The spectrometer has a resolution of 0.15~nm and covers a range of 165--275~nm, leaving it blind to the pump wavelengths in the blue (405--473~nm).
The signal at 202.5~nm wavelength is only detected using the highly sensitive photoemission detector, whereas the other far-UVC signals are recorded both as spectra and with the photoemission detector.

Determining a calibrated measurement of the signal power has not yet been achieved. The significant coupling and propagation losses are accompanied by diffuse scattering of the blue light from the pump at intensities much higher than the far-UVC signal.
Hence, a detector must have both very high sensitivity in the UV while having high suppression in the blue.
By calibrating the spectrometer against a well known KrCl excimer source emitting at 222~nm, the irradiance from the frequency converter is estimated to be 0.17~\textmu{}W/cm$^2$.
This solitary measurement gives no direct value for total output power nor conversion efficiency as it constitutes a very localized intensity in a complex radiation pattern.

A projection screen made from highly UV reflective PTFE (POREX Virtek) is placed in front of the emission region, giving a diffuse reflection of the far-field at the position of the projection screen.
The UV camera is highly selective in favor of the UV light over the visible.
In combination with a band-pass filter centered at 220~nm, the camera setup becomes completely blind to the intense blue pump-laser, imaging only the far-UVC generated from the frequency conversion.

For the SFG measurement, the ECDL at 445~nm wavelength and the free-space laser at 473~nm wavelength are combined in a polarization-maintaining fiber splitter.
By ensuring the polarization is aligned to the same axis at the fiber input, both lasers are coupled to the same waveguide in the frequency converter chip in TE polarization.
To balance the coupling of the three generated UV signals, the angular position of the output spectrometer probe is selected as a compromise because the Cherenkov angle varies with the wavelength.

\bibliography{references}

\section*{Acknowledgements}

We acknowledge support from Innovation Fund Denmark (InnoBooster) and Eureka (Eurostars).
We thank ProxyVision and in particular Rene Lewinski for generously lending us the solar-blind UV Camera and helping to attain the pictures.
Further, we thank Hübner Photonics for providing the 473-nm laser used to demonstrate SFG.

\section*{Author contributions statement}

E.J.S, E.Z.U., P.J. and N.V. conceived the project.
E.J.S., P.T., E.Z.U. and N.V. developed the theory.
E.J.S., P.T. and E.Z.U. designed the structure and fabrication process.
E.J.S., P.T., E.Z.U., S.C., M.A.B., S.T.T. and K.B.G. conducted the experiments.
E.J.S., P.T., E.Z.U., S.T.T. and K.B.G. performed the data analysis.
E.J.S. and P.T. wrote the manuscript with contributions from all authors.
M.A.B. and N.V. supervised the project.

\section*{Data availability}

The datasets generated during the current study are available from the corresponding author on reasonable request.

\section*{Competing interests}

E.J.S. and M.A.B. are members of EMode Photonix.
E.Z.U. and N.V. are shareholders of UVL A/S.
All other authors declare that they do not have any competing interest.




\end{document}